# Three-Dimensional Grain Boundary Spectroscopy in Transparent High Power Ceramic Laser Materials

Mariola O. Ramirez[1*], Jeffrey Wisdom[2], Haifeng Li[3], Yan Lin Aung[4], Joseph Stitt[1], Gary L. Messing[1], V. Dierolf[5], Zhiwen Liu[3], Akio Ikesue[4], Robert L. Byer[2] and Venkatraman Gopalan[1]

[1]*Department of Materials Science and Engineering and Materials Research Institute, Pennsylvania State University, University Park, PA 16802*
[2] *Department of Applied Physics, Ginzton Laboratory, Stanford University, Palo Alto, CA 94304*
[3] *Department of Electrical Engineering, Pennsylvania State University, University Park, PA 16802*
[4] *World-Lab Co., Ltd., Atsuta-ku, Nagoya 456-0023, Japan*
[5] *Department of Physics, 16 Memorial Drive, Bethlehem, PA 18015, USA*

*Corresponding author: mdr20@psu.edu*

**Abstract:** Using confocal Raman and fluorescence spectroscopic imaging in 3-dimensions, we show direct evidence for $Nd^{3+}$-$Nd^{3+}$ interactions across grain boundaries (GBs) in $Nd^{3+}$:YAG laser ceramics. It is clearly shown that $Nd^{3+}$ segregation takes place at GBs leading to self-fluorescence quenching which affects a volume fraction as high as 20%. In addition, we show a clear trend of increasing spatial inhomogeneities in $Nd^{3+}$ concentration when the doping levels exceeds 3 at%, which is not detected by standard spectrometry techniques. These results could point the way to further improvements in what is already an impressive class of ceramic laser materials.

**OCIS codes:** (140.3390) Laser materials processing. (140.5680) Rare earth and transition metal solid-state lasers. (180.6900) Three-dimensional microscopy. (300.6360) Spectroscopy, laser.

## 1. Introduction

The recent development of transparent ceramic laser materials with dense grain boundaries and performance exceeding single crystals is a paradigm shift in the field of laser materials. With solid state processing, it is now possible to achieve an order of magnitude increase in the concentration of active ions over single crystals, window-pane size laser materials, and engineered dopant profiles otherwise not possible in single crystal growth [1-3]. Of particular relevance here is the case of Nd$^{3+}$:YAG ceramics where power scaling up to the kilowatt-level has been demonstrated [4]. On the other hand, the theoretical performance should be even higher; instead a decrease in the laser slope efficiency, a degradation of the beam quality factor and several dynamic instabilities are seen with increased doping, the atomistic and microscopic origin of which are presently not understood [5-8].

Nd$^{3+}$ doped yttrium aluminum garnet; Y$_3$Al$_5$O$_{12}$ (YAG), is currently the most widely employed solid-state laser material for micromachining, medical operations, materials processing and many other industrial and defense applications. Nd$^{3+}$:YAG lasers are based on single crystals, that are restricted to a maximum concentration of 1.4 at. % Nd$^{3+}$, as well as to centimeters in dimensions [9]. This relatively low concentration level leads to a moderate Nd$^{3+}$ pump absorption, strongly limiting the development of lasers of compact size, high power and efficiency. In 1995, Ikesue *et al.* was the first to demonstrate efficient laser action in transparent *polycrystalline* Nd$^{3+}$:YAG with equal or superior optical quality to conventional laser single crystals [10]. Furthermore, up to 9 at.% Nd$^{3+}$ doped YAG ceramics were achieved without any deterioration in the achievable sample size or the optical quality. Since then, because of the significant potential for a new era of laser materials that are *not* single crystals, in addition to the obvious advantages in scaling, processing speed and cost, ceramic lasers gain media has become a highly active area of research worldwide [1,3,11-13].

Despite the extraordinarily rapid progress achieved in transparent laser ceramics, it appears they could be even better! Many open questions remain: How do grain boundaries (GBs) affect the optical response of laser active ions in ceramics? Why is there a variation in the laser performance between samples that are otherwise identical in scattering and doping levels?[14] How are laser slope efficiency, beam quality, and mode stability influenced by grain boundaries? [2,7,8] Otsuka *et al.* postulate the possible presence of saturable absorber-type defects in the GB, but no direct evidence for it exists [8]. Using 3-dimensional confocal Raman and fluorescence spectroscopy, we show evidence for $Nd^{3+}$-segregation at the grain boundaries, subsequent fluorescence quenching and extensive spatial inhomogeneity in dopant concentration, that can explain many of the observed performance behaviors.

## 2. Materials and methods

*Materials synthesis*: Neodymium doped yttrium-aluminum garnet ($Y_3Al_5O_{12}$, $Nd^{3+}$:YAG) transparent ceramics with $Nd^{3+}$ concentrations ranging from 1-5.5 a.% exhibiting nearly the same optical properties as those of a single crystal were fabricated by a solid-state reaction method using high purity powders. High purity $Y_2O_3$, $Al_2O_3$ and $Nd_2O_3$ powders (99.99~99.999% purity) were used as starting materials. These powders were blended with the stoichiometric ratio of YAG including $Nd^{3+}$ of 1.0, 2.3, 3.5, and 5.2 at% and ball milled for 12 h in ethanol with 0.5mass% TEOS (tetraethyl orthosilicate) added as a sintering aid. Then the alcohol solvent was removed by spray drying the milled slurry. Spherical granules (ca. 30~50 µm) having a homogeneous composition were obtained. The spray dried powder was pressed with low pressure into required shapes in a metal mold and then cold isostatically pressed at 98~196 MPa. Transparent $Nd^{3+}$:YAG ceramics were obtained after sintering under vacuum ($1x10^{-3}$Pa) at 1750ºC for 10 hours. Unlike single crystal growth, solid state sintering up to 9 at% $Nd^{3+}$, which is near to 12% solubility for $Nd^{3+}$ ions in YAG, yields homogeneous $Nd^{3+}$ doping without macroscopic gradients. For our studies, all the samples were unetched and surfaces polished to laser quality. No grain boundaries were observable under optical microscope or atomic force microscope.

*Confocal Imaging*: For confocal spectroscopic characterization, room temperature Raman and Fluorescence spectra were recorded by a WITec alpha300 S device using both tunable Argon and He-Ne lasers coupled into a single mode optical fiber. The signal was collected in backscattering geometry with the same objective (100 x magnification, numerical aperture, N.A = 0.9 in air) and focused into a multi-mode fiber (50 µm) that acts as pinhole for confocal microscopy. For the 514 nm excitation wavelength, the laser was focused to a diffraction-limited spot size of about 355 nm. In the axial direction, a conservative estimate of resolution is $2\lambda n/(N.A)2=1270$ nm in air, and approximately 700 nm inside YAG (where $n$=1.84 for YAG (*3*) and N.A=1.84 x 0.9 inside YAG). In order to evaluate the influence of far field contribution to fluorescence, experimental measurements of the doping profile was performed on a junction of a 1 at % doped $Nd^{3+}$:YAG with an undoped single crystal YAG fabricated by diffusion bonding. The resulting measured doping profile shows that the concentration change at the interface (10-90% change) occurs over 800nm, which indicates the 10-90% lateral resolution of the fluorescence imaging in detecting interfaces. Thus, the ~2µm FWHM (25% to 75% change) in fluorescence signal across the grain boundary in Fig 2(d) is well resolved. The far-field contribution was below 10% signal level, observed as a weak "tail" in the fluorescence over 4-5µm. Therefore, for the average grain sizes studied in this work, the far-field contribution is small and can be neglected.

## 3. Results and discussion

The crystal structure of YAG materials belongs to the space group $Ia3d(O_h^{10})$. It contains eight molecular units in the unit cell. Three different sites are available in the lattice. The dodecahedral site *c* with local symmetry $D_2$, is normally occupied by the large $Y^{3+}$ ions surrounded by eight $O^{2-}$ ions, the octahedral site *a* (local symmetry $C_{3i}$) is normally occupied by $Al^{3+}$ surrounded by six $O^{2-}$ ions and the tetrahedral site *d* (local symmetry $S_4$) is occupied by $Al^{3+}$. The $Al^{3+}$ cations occupy eight octahedral sites of $C_{3i}$ symmetry and twelve tetrahedral sites of $S_4$ symmetry. $Nd^{3+}$ ions usually replace $Y^{3+}$ cations placed in twelve dodecahedral sites of $D_2$ symmetry[15]. The large number of atoms in the primitive cell leads to 240 (3 x 80) possible normal modes which can be classified according to the irreducible representation of the $O_h$ group as follows:

$$\Gamma = 18T_{1u} + 3A_{1g} + 8E_g + 14T_{2g} + 5A_{2g} + 5A_{1u} + 5A_{2u} + 10E_u + 14T_{1g} + 16T_{2u} \qquad (1)$$

The 25 modes having symmetries $A_{1g}$, $E_g$ and $T_{2g}$ are Raman active while the 18 having $T_{1u}$ symmetry are IR active. 23 of the 25 Raman active vibrational modes can be observed in the Raman spectrum from 0 to 900 cm$^{-1}$. The Raman spectrum of rare earth doped YAG compounds can be divided into two different parts: the high frequency region (500-900 cm$^{-1}$) and the low frequency region (<500 cm$^{-1}$). The high frequency region accounts to the $\nu_1$ (breathing mode), $\nu_2$ (quadrupolar) and $\nu_4$ molecular internal modes associated with the ($AlO_4$) group, whilst the low frequency region is due to: (i) translational motion of the rare earth ions, (ii) rotational and translational motion of the $AlO_4$ units, and (iii) the $\nu_3$ molecular mode of the $AlO_4$ [16-18]. Since the main goal of our work concerns the state of doping ions in ceramic materials in the vicinity of GBs, we have focused our attention on the low frequency region.

The optically transparent $Nd^{3+}$:YAG samples with $Nd^{3+}$ doping of 1, 2.3, 3.5, and 5.2% at% were prepared by Ikesue *et al.* using solid-state reaction sintering (methods section). We first focus on the 1 at% doped sample. Fig. 1A shows three different Raman spectra in the low frequency region corresponding to three distinct grains in the sample. The relative intensities of the $E_g$ and $T_{2g}$ modes are strongly modified depending on the analyzed grain. On the contrary, the intensity of the $A_{1g}$ lattice symmetric mode peaking at ~370 cm$^{-1}$ remains constant regardless the examined grain. This behavior can be explained in terms of the randomly oriented nature of micro-crystals in polycrystalline ceramics and the Raman scattering intensity calculated from the scattering matrices (M) [16], which predict no intensity change with orientation for the $A_{1g}$ lattice symmetric modes, and changes for the rest. We also do not see any changes for the $A_{1g}$ mode in the region of the grain boundaries. This indicates that no excessive strain is present in these regions. We utilize this property to generate a total *spectroscopic* image of the grain pattern in the sample (Fig. 1B). The changes in Raman intensity for the $E_g$ and $T_{2g}$ modes can be exploited in a variety of ways to visualize the different grains. For instance, a clear distinction between different grains is achieved when the center of mass, defined as $CM = \sum_{i=1}^{n} x_i I_i \Big/ \sum_{i=1}^{n} I_i$, is calculated in the spectral region 360-405 cm$^{-1}$, which changes from grain to grain because $A_{1g}$ mode is constant while the $E_g$ is changed. Fig. 1C shows a different view of the same scanned area obtained by analyzing the integrated intensity of $T_{2g}$ mode at 263 cm$^{-1}$. As observed, not only differences between the grains are observed but also a clear contrast between grains and GBs is manifested. No evidence of GBs was observed in undoped YAG ceramics. Fig.1D shows the evolution of this Raman peak as a function of $Nd^{3+}$ concentration. As the $Nd^{3+}$ content increases, a clear asymmetric broadening and a slight shift to lower wavenumbers is observed, because of the translatory motion of $Nd^{3+}$ ions in combination with the heavy mixing of the translational, rotational and $\nu_3$ mode of the ($AlO_4$) unit. Similar results were obtained for all the Raman modes in the low frequency region. Although the intensity effect observed in Fig.1C is mainly dominated by the orientation of the grains, the obtained contrast at GBs was not observed for the Raman modes in the high frequency region, and hence can be tentatively assigned to different $Nd^{3+}$ distribution between grains and GBs.

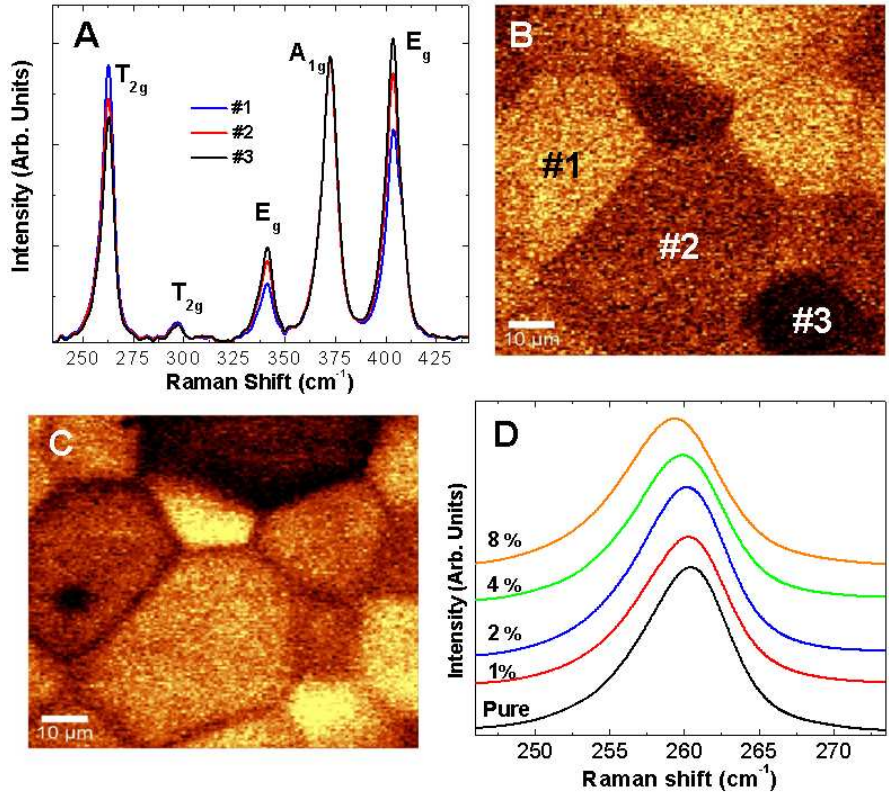

Fig. 1 (A) Detail of the Room Temperature (RT) Raman spectra recorded in confocal geometry at a depth of 10 μm below the surface at three different spatial regions in a transparent 1 at% $Nd^{3+}$ doped YAG ceramic. The excitation wavelength was 488 nm. The symmetry of the different Raman modes has also been included for the sake of clarity. (B) Lateral Raman image (*x-y* scan; 85 x 85 μm) obtained in confocal geometry at a depth of 10 μm below the surface when the center of mass was fixed in the spectral region 360-405 $cm^{-1}$. Note that at every pixel in the Raman image, an entire Raman spectrum was collected (C) Lateral Raman image (*x-y* scan; 85 x 85 μm) obtained by integrating over the 243-273 $cm^{-1}$ spectral region. The scanned area is the same as (B). (D) Raman shift of the $T_{2g}$ mode as a function of $Nd^{3+}$ concentration.

To gain further insight into the $Nd^{3+}$ distribution and activity across a GB, two types of confocal fluorescence experiments were performed: Lateral imaging and Depth Imaging. For the cubic symmetry of the host material and in the absence of excessive stress (as concluded from Raman), a uniform intensity distribution is expected from grain to grain and in the grain boundaries unless variation in $Nd^{3+}$ concentrations and emission quantum efficiency are present. Figure 2A shows the three dimensional (*x* and *y* positions versus intensity) fluorescence image obtained in the lateral geometry by integrating over the whole emission area corresponding to the $^4F_{7/2}$:$^4S_{3/2} \rightarrow ^4I_{9/2}$ and $^4F_{3/2} \rightarrow ^4I_{9/2}$ electronic transitions of $Nd^{3+}$ ions in YAG centered around 800 and 880 nm, respectively. The grain pattern structure in the sample is again reproduced, showing a clear decrease in the total fluorescence intensity when approaching the GB. In particular, we observe an average intensity decrease of ~5 % with respect to the total fluorescence intensity when the emission spectra are recorded at the GBs. Similar images were obtained when the $^4F_{3/2} \rightarrow ^4I_{11/2}$ electronic transition centered at 1060 nm was analyzed. Subtraction of the grain spectrum (~40 μm away from the GB) from the grain boundary spectrum reveals a clear structure at the shoulders of the main optical transition line

(Fig. 2B). Therefore, from Fig. 2B it can be inferred that the contribution of minor $Nd^{3+}$ sites in YAG ceramics differs from grain to GBs. Similar peaks have been observed in samples of higher $Nd^{3+}$ concentration, and have been labeled $M_1$ and $M_2$ spectral satellites [19-21]. It was found, that the $M_1$ satellite, corresponds to the first-order $Nd^{3+}$ pairs (nearest-neighbor (NN)) while $M_2$ is connected with the second order (next-nearest-neighbor (NNN)) pairs, i.e $Nd^{3+}(c)$ - $Nd^{3+}(c)$ pairs in the first and second coordination spheres. With increasing $Nd^{3+}$ concentration, the relative intensities of these satellites increase while those of the isolated ions decrease due to concentration quenching processes activated by direct cross relaxation and migration-assisted energy transfer mechanisms inside the $Nd^{3+}$ pairs. Figure 2C shows the confocal fluorescence image obtained when the contribution of the $Nd^{3+}$ pairs to the total spectrum is analyzed. For such purpose, only the emission area in the satellite peaks, namely, $M_1$ center at 867.6 nm and/or $M_2$ center at 868.7 and 869.9 spectral regions were considered. A notable increase in the emitted satellite intensity is observed across grain boundaries, indicating a higher $Nd^{3+}$ concentration at the grain boundaries. Details of the total emission area corresponding to $M_1$-$Nd^{3+}$ pairs and regular isolated $Nd^{3+}$ ions when crossing a single GB is shown in Fig. 2D. A significant increase of ~ 6% in the contribution of $M_1$ pairs and a corresponding ~4% reduction in the regular isolated $Nd^{3+}$ ions is observed across the GB. This is a clear indication of the increased $Nd^{3+}$ concentration at the GBs, which leads to more pronounced $Nd^{3+}$ segregation and increased $Nd^{3+}$- $Nd^{3+}$ interaction.

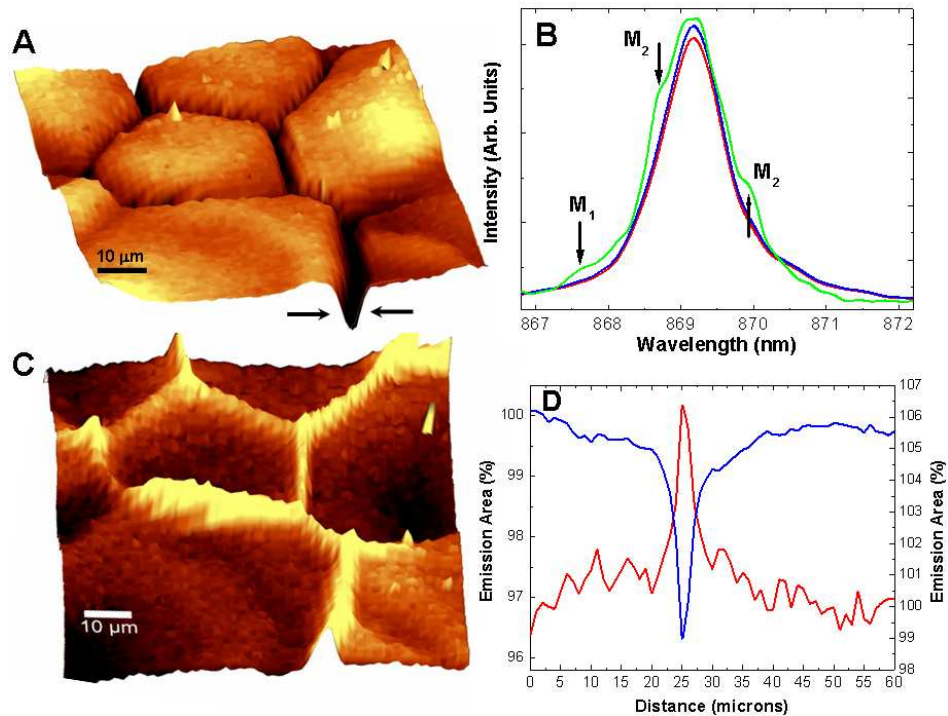

Fig. 2 (A) Three dimensional lateral fluorescence image (*x-y* scan; 85 x 85 μm) obtained in confocal geometry at a depth of 2 μm below the surface by integrating over the whole emission spectrum corresponding to the $^4F_{7/2}$:$^4S_{3/2}\rightarrow{}^4I_{9/2}$ and $^4F_{3/2}\rightarrow{}^4I_{9/2}$ electronic transitions of $Nd^{3+}$ ions YAG. The $Nd^{3+}$ concentration was 1 at% (B) Green line: detail of the most intense line in the emission spectrum obtained subtracting the emission spectra from grains and GBs. For the illustration purposes, it has been multiplied by 20. Blue and red lines: emission spectra collected at grain and GB, respectively. (C) Confocal fluorescence image obtained when the contribution of the $Nd^{3+}$ pairs to the total spectrum is analyzed. Note that only the emission area in the 867.7-868.5 nm ($M_1$-$Nd^{3+}$ pairs) and/or 869.8-870.6 ($M_2$-$Nd^{3+}$ pairs) spectral regions was considered. (D) Relative changes obtained in the total intensity of both, $M_1$-$Nd^{3+}$ pairs (red line, right y-axis) and regular isolated $Nd^{3+}$ ions (blue line, left y-axis) when crossing a single grain boundary.

Thus, we can attribute the reduction in fluorescence quantum efficiency at the GBs to a more pronounced non-radiative decay channel. From measurements of the average quantum efficiency as a function of $Nd^{3+}$ concentration, we can obtain a calibration curve for quantum efficiency versus concentration. Using it, we can conclude that the $Nd^{3+}$ concentration is 0.1 at% higher at the GBs. This is also consistent with atomistic modeling studies in $Nd^{3+}$:YAG that predict a higher $Nd^{3+}$ content was predicted at GBs [22]. Besides the decrease of emission quantum efficiency, the increased concentration and the resulting stronger correlation leads to an average decrease of ~10µs (or ~5%) in the fluorescence lifetime at the GBs in this 1 at% $Nd^{3+}$-doped sample, as estimated from a direct lifetime versus concentration measurement performed independently. Two main reasons can account for the observed $Nd^{3+}$ segregation at GBs, namely solute drag effects due to the increased atomic mass and ionic radius of $Nd^{3+}$ ion with respect to either $Y^{3+}$ or $Al^{3+}$ [23], and the influence of surface morphologies of grain boundaries on the relative concentration of $Nd^{3+}$ ions in YAG [22]. Our results illustrate the exceptional sensitivity of our optical measurement technique. Concentration changes of this order were not detectable by other standard techniques such as Energy Dispersive Spectrometry (EDS), secondary ion mass spectroscopy (SIMS) and Electron Probe Microanalysis (EPMA).

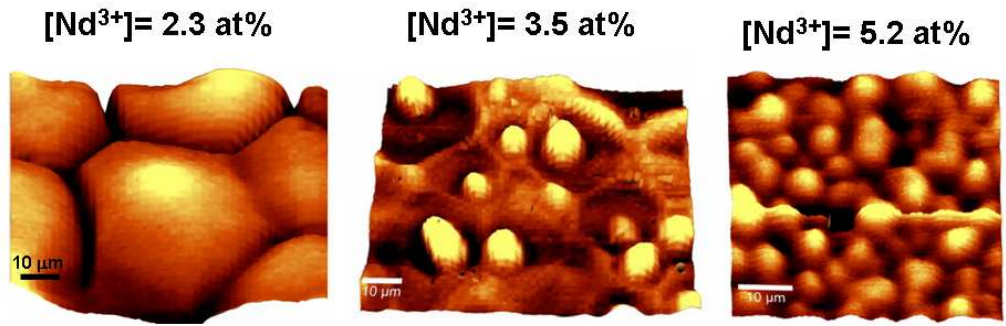

**Figure 3** Three dimensional lateral fluorescence image (*x-y* scan; 85 x 85 µm) obtained in confocal geometry at a depth of 2 µm below the surface by integrating over the whole emission corresponding to the $^4F_{7/2}$:$^4S_{3/2} \rightarrow {}^4I_{9/2}$ and $^4F_{3/2} \rightarrow {}^4I_{9/2}$ electronic transitions of $Nd^{3+}$ ions YAG for three different $Nd^{3+}$ concentrations: 2.3 at% (left side), 3.5 at% (center) and 5.2 at% (right side).

Similar studies were performed on highly doped $Nd^{3+}$:YAG ceramics. Figure 3 shows the lateral confocal images obtained for different $Nd^{3+}$ concentrations ranging from 2 at% up to 5.5 at% Besides a decrease in the grain size with $Nd^{3+}$ concentration as previously reported in other works [6], a gradual decrease in the optical quality was observed for $Nd^{3+}$ concentrations exceeding 3 at%. Our studies show that not only is the $Nd^{3+}$ segregation observed at the GBs, but also non-uniform dopant distribution within the grains (~2% variation, or 0.07 at%, for the 3.5 at% sample and ~4% variation, or 0.21 at%, for the 5.2 at% sample) was observed in different areas in the sample. The yellow contrast regions observed in Fig. 3 indicate regions of higher $Nd^{3+}$ content. A similar study of the commercial Konoshima Inc $Nd^{3+}$:YAG material with 4 at% doping reveals similar spatial inhomogeneities (0.7%, or an upper limit variation of 0.028 $Nd^{3+}$ at%). Because of the efficient energy transfer process present in $Nd^{3+}$:YAG system, the observed spatial inhomogeneities in doping leads to locally enhanced energy transfer efficiencies among $Nd^{3+}$ ions and may explain the observed laser instabilities. More specifically, we believe these spatial heterogeneities can act as saturable-absorber type defects as postulated in literature without direct evidence [8], that can explain the dynamical instabilities and the strong beam degradation observed in highly doped $Nd^{3+}$:YAG ceramics. Figure 4 shows a 3-dimensional spatial reconstruction of GBs in an 85 μm³ volume obtained from the emission spectra of $Nd^{3+}$ ions in YAG. Because self-fluorescence quenching across GB spreads over a couple of microns, nearly a quarter of the ceramic laser gain media (~20% volume) presents additional optical losses.

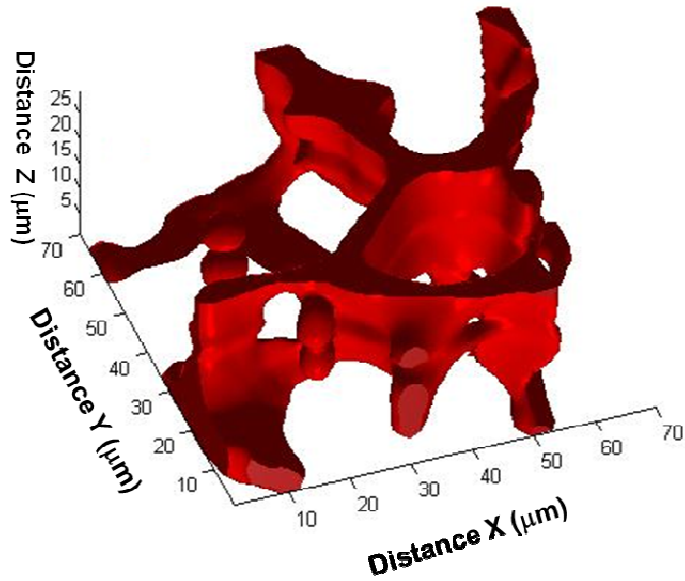

Fig. 4 Partially bottom view at a fixed elevation of the 3D isosurface spatial reconstruction of GB surfaces obtained from the whole emission spectra area of $Nd^{3+}$ ions in YAG. $Nd^{3+}$ concentration was 1 at%.

## 4. Summary and Conclusions

In summary, we have presented a non-invasive, sensitive and powerful method of 3D visualization of grains and grain boundaries in unetched transparent $Nd^{3+}$:YAG ceramics based on spatially resolved confocal spectroscopy. Our results show $Nd^{3+}$ segregation and a decrease in the total fluorescence quantum efficiency of $Nd^{3+}$ ions at GBs. Additionally, an increasingly non uniform $Nd^{3+}$ distribution is found with $Nd^{3+}$ doping level exceeding 3 at%. The observed $Nd^{3+}$ segregation and spatial variations (0.1-0.2 at%) would give rise to spatial refractive index changes on the order of $10^{-5}$ on the grain scale [24], and hence, additional scattering losses, phase front distortions of the laser beam as well as slight perturbations of the $TEM_{00}$ mode by populating higher order modes. For example, spatial index variations of $10^{-5}$ can lead to GHz shifts in laser tuning frequency over a 10cm rod length, which is 2-3 orders of magnitude higher than typical laser linewidths (MHz), thus leading to temporal instability [25,26]. This study indicates that grain sizes on the order of 1-2 microns are preferred for laser experiments since the microscopic dopant gradient due to GBs segregation (~2 μm) is overcome. Interestingly, the commercial Konoshima Nd:YAG material, which has the highest laser performances reported, has a grain size smaller than 2μm, which would support the conclusion of this study. Though the study focuses on $Nd^{3+}$:YAG, the proposed study can be broadly applied to other ceramic laser materials.

### Acknowledgments

We would like to acknowledge support from the Center for Optical Technologies at Pennsylvania State University, and VLOC Inc.